\documentclass[10pt,conference]{IEEEtran}
\IEEEoverridecommandlockouts
\usepackage{cite}
\usepackage{amsmath,amssymb,amsfonts}
\usepackage{algorithmic}
\usepackage{graphicx}
\usepackage{textcomp}
\usepackage{xcolor}
\usepackage{algorithm}
\usepackage{subcaption}
\usepackage{soul}
\usepackage{hyperref}

\def\BibTeX{{\rm B\kern-.05em{\sc i\kern-.025em b}\kern-.08em
    T\kern-.1667em\lower.7ex\hbox{E}\kern-.125emX}}

\begin{document}

\title{Routing in Quantum Repeater Networks\\with Mixed Efficiency Figures\\

}

\author{\IEEEauthorblockN{Vinay Kumar\IEEEauthorrefmark{1}\IEEEauthorrefmark{2}, 
Claudio Cicconetti\IEEEauthorrefmark{2},
Marco Conti\IEEEauthorrefmark{2}, and
Andrea Passarella\IEEEauthorrefmark{2}} \\
\IEEEauthorblockA{\IEEEauthorrefmark{1}Department of Information Engineering, University of Pisa, Pisa, Italy}
\IEEEauthorblockA{\IEEEauthorrefmark{2}Institute for Informatics and Telematics (IIT), National Research Council (CNR), Pisa, Italy}}

\maketitle

\begin{abstract}

This study explores an approach to routing in quantum networks, which targets practical scenarios for quantum networks, mirroring real-world classical networks. By addressing practical constraints, we examine the impact of heterogeneous nodes with mixed efficiency figures on quantum network performance. In particular, we focus on some key parameters in an operational quantum network such as the fraction of nodes with a higher efficiency (called high-quality), path establishment order, end-to-end fidelity, i.e., a measure of the quality of the end-to-end entanglement established. Our simulations show that incorporating knowledge of node quality not only helps boost the fidelity of some of the routing paths but also reduces the number of blocked paths in the quantum network. The study also highlights the critical role of the fraction of high-quality nodes in end-to-end fidelity and explores the trade-offs between upgrading all nodes to high quality or retaining a subset of lower-quality nodes. 
\end{abstract}


\bigskip
\noindent\textbf{Note.} This is the author accepted manuscript.
\par
\noindent \textbf{Please cite as}:
V. Kumar, C. Cicconetti, M. Conti and A. Passarella, 
``Routing in Quantum Repeater Networks with Mixed Efficiency Figures,'' 
2024 IEEE Future Networks World Forum (FNWF), Dubai, United Arab Emirates, 2024, pp. 198--203. DOI: \href{https://doi.org/10.1109/FNWF63303.2024.11028804}{10.1109/FNWF63303.2024.11028804}

\section{Introduction}\label{sec:intro}


Leveraging the principles of quantum mechanics, quantum networks aim to provide more secure communication and improved computational capabilities compared to classical networks \cite{kim08}. A quantum network is a set of \textit{quantum repeaters} or nodes connected by quantum and classical links \cite{brigel98}. 
Nodes on a quantum network can be categorized into three generations based on their operational errors and the techniques used to mitigate the signal loss~\cite{munro15, murli16}. Since the initial development of quantum networks is expected to operate first generation (1G) of nodes, which are simplest and, hence, easier to engineer and manufacture at scale, we focus on them in this paper.

\textit{Quantum entanglement} is a phenomenon where the states of two or more qubits are correlated to each other independent of the spatial distance between them. For a system of two particles, it is called an 'EPR pair' \cite{epr35}. 
In reality, an entanglement attempt across the quantum channel is a probability that depends on the loss in the quantum channel \cite{pirandola16, murli16}. This often leads to the failure of an entanglement attempt which is referred to as a `link failure' in quantum network terminology.

\begin{figure}[tb]
    \includegraphics[width=\columnwidth ]{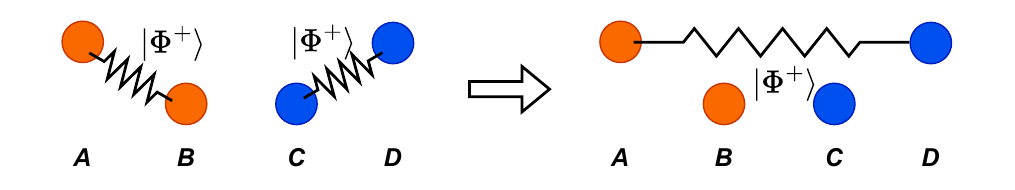}
    \caption{Entanglement swapping procedure}
    \label{fig:swap}
    \vspace{-0.5cm}
\end{figure}

\textit{Entanglement swapping} is used to establish entanglement between qubits that have never directly interacted with each other \cite{zuk93} as shown in figure \ref{fig:swap}. 
Two individual entangled pairs of qubits i.e., AB and CD results in an entangled pair of qubits AD over long-distance under entanglement swapping procedure. This procedure consists of simultaneously measurement of qubits B and C, communicating result to at least one of A and D, and informing the other that the measurement is done. Upon receiving the communication, relevant quantum gates are applied to A and D to complete the entanglement swapping procedure. This enables the creation of long-distance entangled connections by `chaining' together shorter entangled links \cite{san11}. 

A metric \textit{fidelity} is used to quantify the similarity between two states. It takes a value between 0 and 1, where 1 indicates that the states are
identical, and 0 implies that they are completely different. Depending on the application a minimum fidelity is required, which should be always above 0.5 for practical purposes.

One of the fundamental problems of quantum networks is \textit{entanglement routing} \cite{gyon22}. 
Similar to classical networks, a quantum network needs a routing protocol to find the most efficient path between source and destination nodes. However, due to the fragility of quantum states and the impossibility to amplify/clone them, quantum routing has unique challenges associated and continues to be an active area of research.

This paper explores the impact of heterogeneous nodes on entanglement routing paths and demonstrates a technique to boost the performance of quantum networks in simple, yet practical, settings.
The ultimate goal of quantum networking is to deploy a \textit{quantum internet} merged with classical networks to enable secure communications, distributed quantum computing, and other quantum applications \cite{simon17, weh18}. In this context, our study presupposes a network architecture wherein end nodes are situated within access networks, which are subsequently interconnected through a core transport network. The devices in this core transport network exhibit varying degrees of efficiency due to an incremental deployment of nodes that become increasingly more stable and sophisticated over time due to technology advances. This is a key novelty of our approach: to the best of our knowledge, this is the first study that specifically includes this aspect in the system model, while prior works assume that all nodes are identical, even though link quality may vary due to distance-based attenuation.

\section{Related Works}\label{sec:works}
Meter et al. \cite{meter12} examined quantum networks for path selection and proposed link cost-based routing metric. 
The cost associated with a path for routing is calculated in an additive manner using proposed metrics, and the path is determined using Dijkstra's shortest-path algorithm. 

Pant et al. \cite{pant2019routing} developed a protocol to achieve higher entanglement rates by using multiple paths while considering inefficient links in the quantum networks. A time slot smaller than the qubit decoherence time is divided into external and internal phases to establish a routing path from a source node to its destination. In the external phase, dedicated quantum memories of neighbouring nodes attempt to establish a link between the nodes via a shared entangled pair. On the other hand, the internal phase consists of establishing an internal link within the node's quantum memory to complete the routing path between the source and destination nodes. 

As an extension to this, S. Shi et al. \cite{shi2020} proposed an algorithm to increase the number of successful long-distance entanglements. Instead of two phases, the time slot consists of four phases to establish a routing path between sources and destinations. In addition to the external and internal phases, the other two phases comprise of receiving information regarding current source-destination pairs required to establish the connection and sharing classical information between neighbouring nodes to learn their respective link states. 

C Li. et al. \cite{cli21} proposed a generalized scheme to Pant et al. \cite{pant2019routing} to handle multiple requests in addition to a multi-path approach while ensuring fidelity using \textit{entanglement purification}, a technique where some entangled qubits are consumed to produce a single higher-quality pair. Purification can be a viable solution to increase the fidelity of entangled qubits and we leave the assessment on how to incorporate it in our contribution for future study.




We extend the findings in prior works by studying the case of a quantum network with heterogeneous nodes, in terms of the amount of efficiency of a quantum repeater in applying quantum gates in the entanglement swapping procedure. Furthermore, in the performance evaluation, we consider an engineered quantum network, where \textit{server} nodes are topologically separated from \textit{client} nodes by a transport network of quantum repeaters, which mimics practical deployments in the classical Internet.

\section{Network and motivation}\label{sec:network_and_motivation}

\subsection{System Design}\label{ssec:sys_des}
\begin{figure}[tb]
    \includegraphics[width=\columnwidth ]{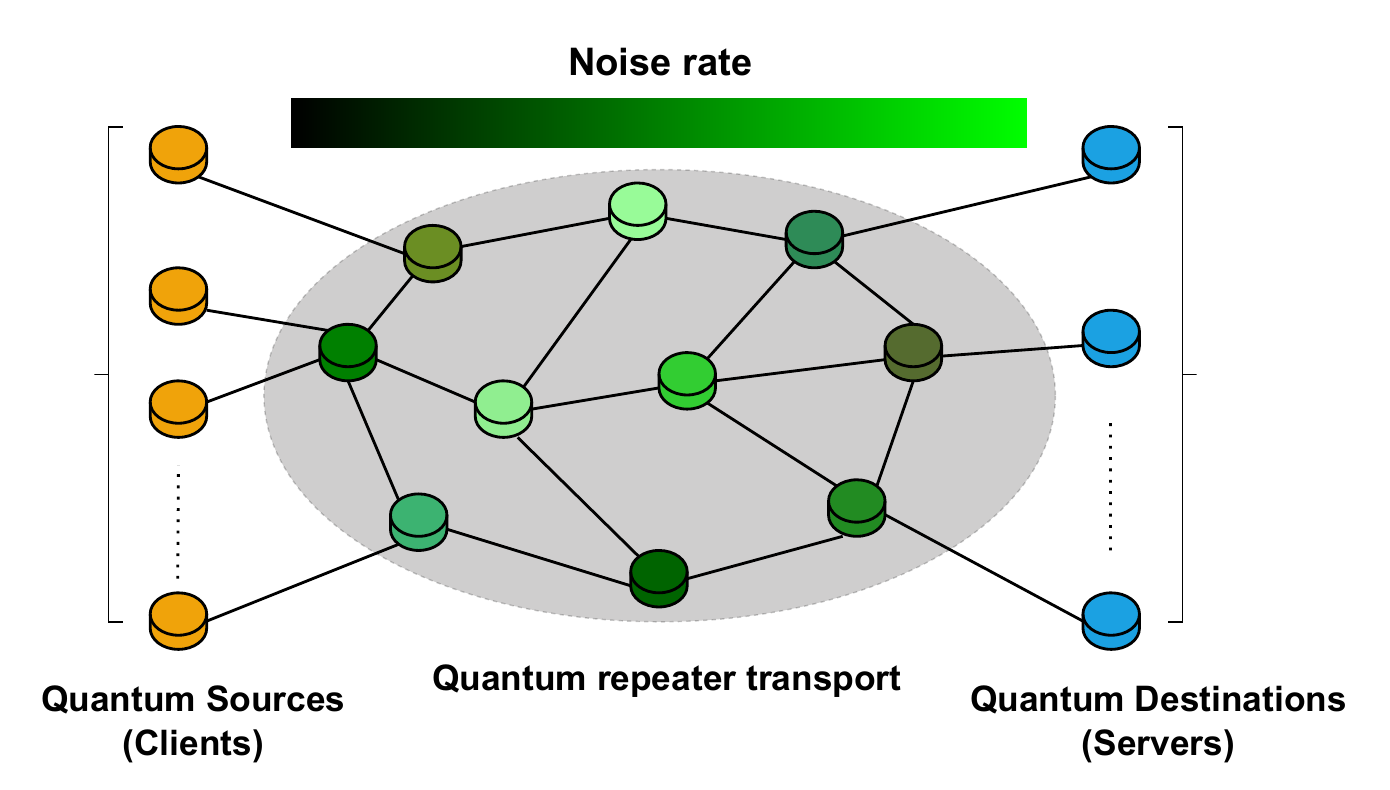}
    \caption{A general schematic of a quantum network}
    \label{fig:schematic}
    \vspace{-0.5cm}
\end{figure}

We illustrate the quantum network by constructing an undirected graph as shown in figure~\ref{fig:schematic}, where the nodes on this graphical representation symbolize the presence of nodes on the quantum network. The edges of the graph represent quantum network links which are typically realized using fiber optic cables having classical and quantum channels capable of transferring quantum states and classical information. 
The network consists of \textit{transport network} (shown in green shades) that defines the network topology and is comprised of nodes with a range of efficiencies associated with them. From a technology readiness perspective, the quantum network is still a novel concept in its early stages of development. According to the Technology Readiness Level (TRL) scale, the current status of quantum networks would likely be in the range of TRL 3-4, where basic proof-of-concept experiments are being conducted, and small-scale testbeds (typically with a few nodes) are emerging experimentally.

In general, several rounds of entanglement attempts are performed per time slot to overcome the `link failure' issue discussed in section~\ref{sec:intro}.
However, this is a very resource-intensive approach and requires a high EPR generation rate and a large number of qubits. Recent results have shown that it is possible to generate entanglement deterministically within a time slot \cite{humphreys18}. This can also be achieved by a connection-oriented protocol which guarantees the generation rate between source and destination nodes~\cite{jli22}. 

As discussed in section~\ref{sec:intro}, entanglement swapping procedures rely on quantum state measurements to distribute entanglement across a quantum network. However, these measurements are inherently imperfect due to the finite efficiency of the measurement devices used. The impact of this imperfection is quantitatively analyzed within quantum repeaters or nodes, as follows
\cite{helstrom69, dur99}:
\begin{equation}\label{eq:P0}
P_0 = \eta |0 \rangle \langle 0| + (1 - \eta) |1 \rangle \langle 1|
\end{equation}
\begin{equation}\label{eq:P1}
P_1 = \eta |1 \rangle \langle 1| + (1 - \eta) |0 \rangle \langle 0|
\end{equation}
where $\eta$ is the efficiency of the qubit measurement in the basis states.
Consider a scenario where the correct measurement basis for a qubit is $|1 \rangle \langle 1|$. Then, during entanglement swapping in quantum repeaters, the measurement is characterized by the Eq.~(\ref{eq:P1}). For an ideal measurement device in quantum repeater $\eta = 1$, the measurement results are entirely reliable. However, if $0 \leq \eta < 1$, there exists a non-zero probability of $1-\eta$ that the results will be incorrect. The same probability of error applies to measurement in the $|0 \rangle \langle 0|$ basis described by Eq.~(\ref{eq:P0}).

Routing is assumed to be a centralized process in this work. A linear routing path between sources and destinations is visualized involving N entangled qubit pairs.
The entanglement swapping procedure is performed to establish an entanglement connection between source (A) and destination (B) shown as discussed in section~\ref{sec:intro}. 



For a linear path consisting of repeater nodes belonging to $G$ classes of efficiencies, representing different vendors of devices, or different models for the same vendor. The end-to-end fidelity for such a linear path is given by \cite{dur99}:

\begin{equation}\label{eq:2}
\begin{split}
F_N = \frac{1}{4} \biggl\{ 1+3 \prod_{g=1}^G \left[ \left( \frac{4 {\eta_g}^2 -1}{3} \right)^{N_g} \left( \frac{4 F -1}{3} \right)^{N_g} \right] \\
\left[\frac{4
F -1}{3}\right] \biggr\},
\end{split}
\end{equation}
where
$\eta_g$ is the efficiency of the $g^{th}$ class of node,
$N_g$ is the number of nodes of the $g^{th}$  class in the linear path, and F is the initial fidelity of the general EPR pairs.


Formally, we define the \ul{Quantum Routing With Mixed Efficiency Problem as follows}:
\textit{Given a quantum network $G(V,E)$ where the repeaters belong to $G$ with characteristic $\eta_g$ and generate EPR pairs with fidelity $F$, a set of $P$ source-destination pairs, and a minimum fidelity threshold $\bar{F}$, find a set of non-overlapping paths such that the end-to-end fidelity computed according to Eq.~(\ref{eq:2}) is above the threshold $\bar{F}$.}

Depending on the network topology and the other requirements, it is possible that not all the source-destination pairs can be assigned a path.
When that happens we say that the source-destination pair was \textit{blocked}, borrowing the term from call admission control schemes terminology.
Of course, during the system operation, it is desirable that blocking happens as sparingly as possible.
In principle, it is possible to convert the problem above to a minimization problem, where the objective is, indeed, to minimize the number of blocked source-destination pairs, which would be very computationally challenging to solve due to the integrity of variables and the structure of constraints following Eq.~(\ref{eq:2}).
Instead, in the following we adopt a simple and practical path selection procedure, described through the pseudo-code in Algorithm~\ref{alg:shortest-path-entanglement}, which is sufficient for our purposes, i.e., assessing how some of the key system factors affect the performance.
Since the procedure is greedy, i.e., it tries to assign a path to source-destination pairs following their ordering and without backtracking previous decisions, the ordering affects significantly the output of the procedure.
This bias will be removed in the analysis in section~\ref{sec:performance} by repeating multiple times the selection with random ordering, which allows us to gather performance in an average case.
The path selection algorithm can be easily extended to include different values of the initial fidelity $F$ of nodes, and a different fidelity threshold $\bar{F}$ for each source-destination pair, but we assume network-wide values for better readability and consistency with the performance evaluation below.

\begin{algorithm}
\caption{Vertex-weighted Dijkstra's algorithm}
\label{alg:shortest-path-entanglement}

\textbf{Input}:
Network as a graph $G(V,E)$, efficiency of all the nodes $\{ \eta_i \}, i \in V$, source-destination pairs $\{ s_k, d_k \}, k = 1..P$, where $P$ is the number of pairs, fidelity threshold $\bar{F}$, weight mapping function $f(\cdot) : \eta \rightarrow w$

\textbf{Output}:
List of end-to-end entangled paths and associated fidelity $\{ \pi_j, F_j \}$ and the number of source-destination pairs for which no path was allocated ($B$)

\begin{algorithmic}[1]
\STATE Paths = $\emptyset$
\STATE B = 0
\STATE Shuffle source-destination pairs
\STATE Derive weighted graph $G'(V',E')$ where $v \in V$ is assigned a weight based on its $\eta$ via $f(\cdot)$ and $E'=E$
\FOR {each pair $\{ s, d \}$}
\STATE Find the shortest path $\pi$ between $s$ and $d$ in $G'$
\STATE allocated=False
\IF {path $\pi$ is found}
\STATE Compute fidelity $F$ of the path $\pi$
\IF {$F \ge \bar{F}$}
\STATE Paths += $(\pi, F)$
\STATE Remove edges in $\pi$ from $E'$
\STATE allocated=True
\ENDIF
\ENDIF
\IF {allocated is False}
\STATE B++
\ENDIF
\ENDFOR
\STATE return \{ Paths, B \}
\end{algorithmic}
\vspace{-0.3em}
\end{algorithm}

\textit{By appropriately tuning $f(\cdot)$ it is possible to obtain a different path selection behavior that is more or less sensitive to the nodes' efficiency figures.} 
\begin{figure*}[tb]
    \includegraphics[width=2\columnwidth]{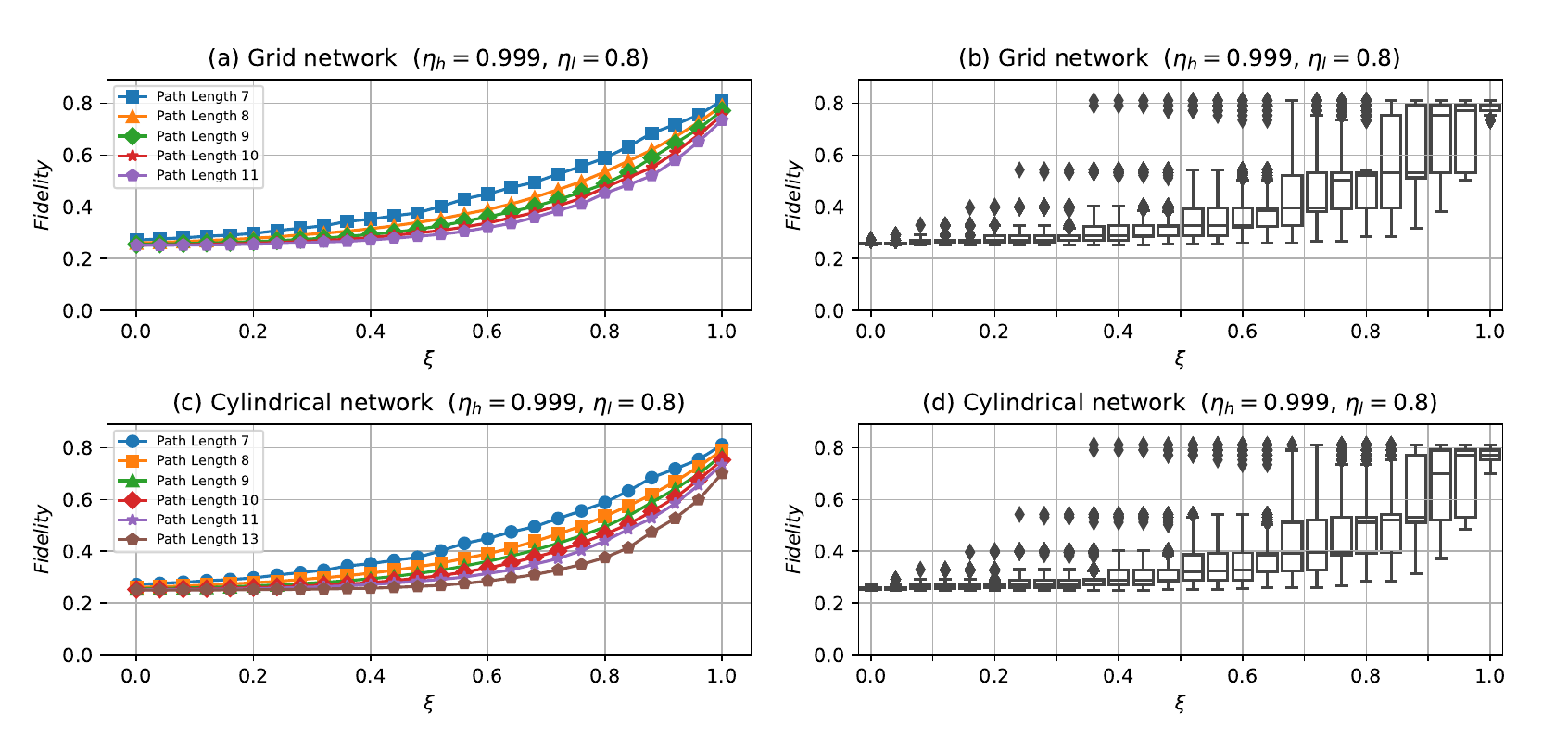}
    \vspace{-0.4cm}
    \caption{a) \& b) Fidelity vs $\xi$ for Grid network, c) \& d) Fidelity vs. $\xi$ for Cylindrical network}
    \label{fig:fid_vs_hq}
    \vspace{-0.5cm}
\end{figure*}
\section{Performance Evaluation}\label{sec:performance}
In this section, we first discuss the methodology adopted in the quantum network simulations (\ref{ssec:method}), and then we display and discuss the results (\ref{ssec:results}).

\subsection{Simulation tool and Methodology}\label{ssec:method}
We constructed a specialized Python simulator for the system outlined in Section~\ref{ssec:sys_des}. The code-base for all simulations is freely available on GitHub in the repository $vk9696/quantum-routing1$. This simulator models a quantum network that follows a time-slotted model, analogous to methodologies described in prior research \cite{pant2019routing, shi2020, cli21}. 

We consider a quantum network with either a regular $n \times n$ grid or a cylindrical grid, where the latter adds connections between the top and bottom nodes. This eliminates edge effects and ensures a path between any source and destination. The regular grid mirrors classical mesh networks, commonly seen in metropolitan area networks (MANs), while the cylindrical grid resembles backbone networks in the classical internet, providing resilience and load balancing, similar to SONET/SDH ring topologies.

To analyze and understand the problem, we consider a simple network model which has quantum repeaters of only two classes, while leaving the general setting for future works. Therefore, depending on the efficiency, a node is assumed to be categorized as either High Quality (HQ) or Low Quality (LQ). We define the parameter $\xi$ as the ratio of the number of HQ nodes to total nodes in the transport network: $\xi = 0$ ($\xi = 1$) means that all the nodes are LQ (HQ).
Additionally, we use $\theta$, called path establishment order, to indicate the position of the requested pair (source, destination) after shuffling in Algorithm~\ref{alg:shortest-path-entanglement}.
A request that has a low $\theta$ is more likely to be allocated than one with a high $\theta$, in the same time slot.



\subsection{Results}\label{ssec:results}
In this section, we present the results for the system introduced in section~\ref{ssec:method}. Throughout this study, a $5 \times 5$ core grid is used in both grid and cylindrical networks. A total of five random combinations of sources and destination nodes are considered in addition to 100 random assignations of HQ and LQ nodes to the transport network. Each combination has been repeated 100 times, with random ordering of the source-destination pairs, as mentioned in section~\ref{ssec:sys_des}. We assume $F = 0.975$. 
Our analysis consists of four macro scenarios, which tackle four different aspects: quantum network topology, low-quality efficiency sensitivity, efficiency awareness, blocking probability.

\subsubsection{Quantum Network Topology Study}
In this aspect, the methodology described above is applied to both grid and cylindrical quantum networks.

We set $\eta_h = 0.999$ and $\eta_l = 0.8$ and we use $f(\cdot) = 1$ and $\bar{F} = 0$, i.e., the path selection is independent of the efficiency figure of nodes and their fidelity.

Due to topology, the path length of grid vs. cylindrical networks is between 7 and 11 hops (or 13 only for the cylindrical case), with an average path length of 8.61 (grid) and 9 (cylindrical). As already introduced, paths are never blocked in a cylindrical network, while the blocking probability in the grid network, which has fewer resources (links), is found to be 0.28. This is a direct implication of more resources (i.e., additional links) in cylindrical networks. Importantly, it should be noted that the paths which are blocked in the grid network are redundant for any practical purposes of establishing end-to-end entanglement, and therefore, these are not considered in figure~\ref{fig:fid_vs_hq}.

In figures~\ref{fig:fid_vs_hq}a and \ref{fig:fid_vs_hq}c we report the average fidelity when  $\xi$ increases from 0 to 1, i.e., when the network has an increasing fraction of HQ nodes.
As expected, all the curves increase with $\xi$, and shorter paths have a higher fidelity. We note that the presence of the 13-hops paths in the cylindrical case yields a slightly decreased  average fidelity (0.3896 vs.\ 0.3972 for the grid case -- not shown in the plots).

The fidelity is also reported in figures~\ref{fig:fid_vs_hq}b and \ref{fig:fid_vs_hq}d as box plots. As can be seen, for $\xi < 0.32$ in the grid network and for $\xi < 0.36$ in the cylindrical network, fidelity is almost zero apart from outliers. Hence, it is of no use to upgrade 32\% in the grid network and 36\% in the cylindrical network of LQ nodes to the more expensive/sophisticated HQ nodes. The LQ nodes are observed to be bottlenecks in the performance for both grid and cylindrical networks. Fidelity is barely 0.5 even if $\xi$ is 0.8 in the grid network and 0.84 in the cylindrical network. However, a significant jump in fidelity, of about 50\%, is observed if one more LQ node is replaced with an HQ node for both grid and cylindrical networks. In conclusion, to have a useful impact on network performance from switching some of the LQ nodes to HQ nodes in a quantum network, $\xi$ should be very close to 1, i.e., in our simulations at least 0.84 for the grid network and 0.88 for the cylindrical network.

\subsubsection{Low-quality efficiency sensitivity study}
In this aspect, we study the sensitivity of $\eta_l$ for a constant $\eta_h$ in the cylindrical core-based quantum network: while keeping $\eta_h = 0.999$ we sweep $\eta_l \in \{0.99, 0.8\}$.
Like in the previous scenario, we assume $f(\cdot) = 1$ and $\bar{F} = 0$.

\begin{figure}[tb]
    \includegraphics[width=\columnwidth]{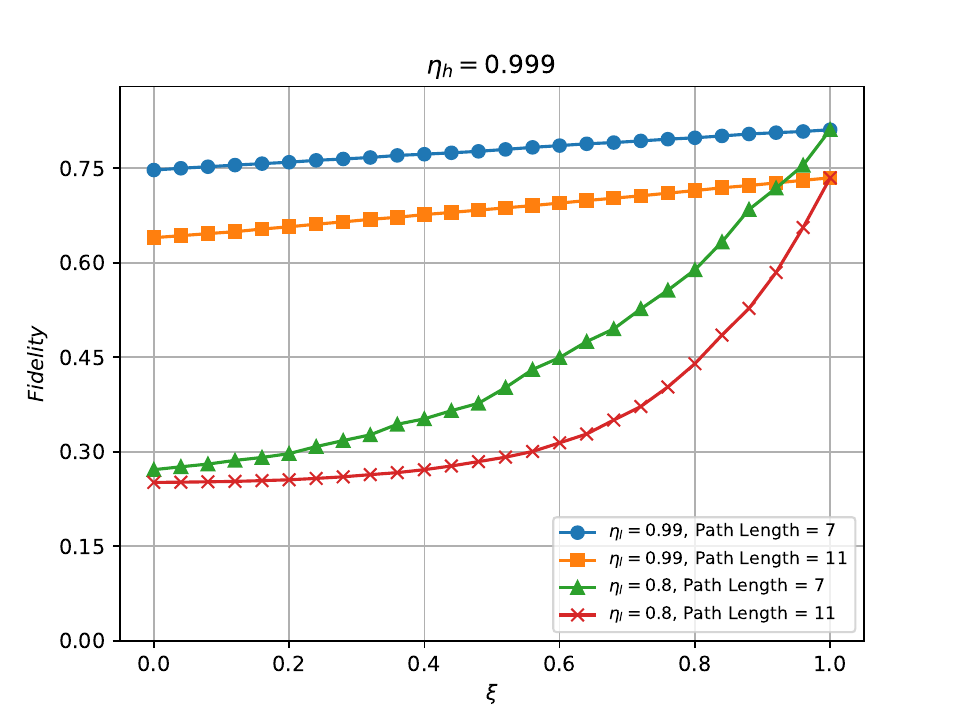}
    \vspace{-0.6cm}
    \caption{Fidelity vs. $\xi$}
    \label{fig:sub_fid_vs_hq}
    \vspace{-0.5cm}
\end{figure}

We show the fidelity in figure~\ref{fig:sub_fid_vs_hq}, only for paths of length 7 and 11 for better readability. As can be seen, a relatively higher $\eta_l$ is able to achieve considerably higher fidelities even if path length and $\xi$ are low as shown by blue and orange squared points. Conversely, if $\eta_h$ is not high enough then the network performance cannot be improved irrespective of path lengths and $\xi$ involved as shown by green and red triangle points. In conclusion, for a network consisting of LQ nodes with a high efficiency figure, replacing some of the nodes with HQ ones is not effective.



\subsubsection{Efficiency Awareness Study}\label{sssec:noise study}
In this aspect, we study the efficiency awareness approach against the standard shortest-path approach in the cylindrical-core based quantum network. We keep $\bar{F}=0$ but consider two possible mapping functions as follows:
\begin{equation}
f_{\mathrm{shortest-path}} = 1 \ \forall \ \eta
\end{equation}
\begin{equation}
f_{\mathrm{efficiency-aware}} = 
\begin{cases} 
100 & \text{if } \eta = \eta_l \\
1 & \text{if } \eta = \eta_h 
\end{cases}
\end{equation}
\begin{figure}[tb]
    \includegraphics[width=\columnwidth]{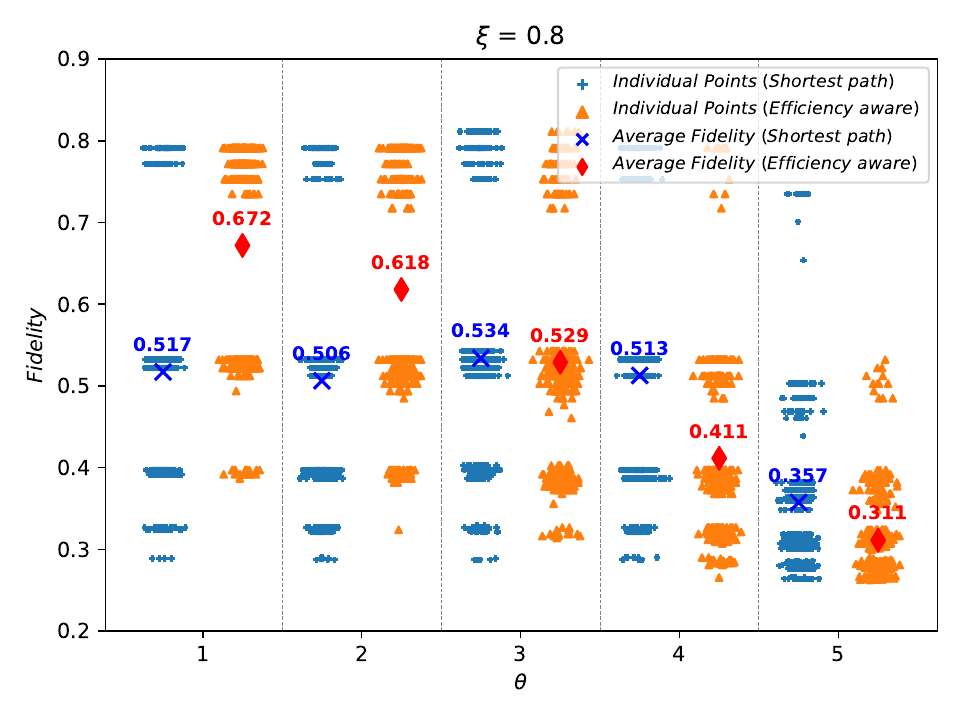}
    \vspace{-0.6cm}
    \caption{Fidelity vs. $\theta$ for $\xi$ = 0.8}
    \label{fig:fid_theta}
    \vspace{-0.5cm}
\end{figure}
In the standard shortest-path selection approach, the efficiency figures of quantum repeaters are not taken into account, leading to uniform weighting of all nodes. Conversely, in the efficiency-aware approach, nodes of lower quality are assigned disproportionately higher weights compared to those of higher quality.

The figure~\ref{fig:fid_theta} shows a scatter plot of the fidelity values for $\xi$ = 0.8 for the two mapping functions above. We consider the standard shortest-path case first. The study shows that for $\xi > 0.2$, the fidelity of the first four paths is found to be closer to each other with a considerably lower fidelity for the last path. This is clearly evident from figure~\ref{fig:fid_theta}. 
i.e., $\xi$ = 0.8, where the first four paths have an average fidelity of approximately 
0.51 while the last path has a fidelity much lower i.e., 
0.31 

In the efficiency-aware scenario, there is a noticeable linear decrease in fidelity with increasing values of $\theta$. This trend is attributable to the consumption of high-quality (HQ) nodes by the initial paths, which subsequently leaves a predominance of lower-quality (LQ) nodes for later paths. Figure~\ref{fig:fid_theta} illustrates significant improvements in higher fidelities for the efficiency-aware approach (depicted in orange) over shortest-path approach (in blue), particularly at the initial stages of $\theta$. The contrast is starkly evident as there is a substantial increase in the number of paths served at higher fidelities for the efficiency-aware method compared to shortest-path approach for $\theta = \{1, 2 \}$. However, for $\theta$ = 3, the average fidelity of the paths served by both approaches converges. Notably, for $\theta$ values of 4 and 5, shortest-path approach outperforms the efficiency-aware method. This pattern indicates that the efficiency-aware approach prioritizes enhancing the fidelity of initial paths, thereby affecting the overall distribution of fidelity in subsequent paths.

For the scenarios where $\xi = 0$ and $\xi = 1$, the fidelity distribution remains consistent across all values of $\theta$ for both shortest-path and efficiency-aware approaches, as anticipated. This consistency arises because, at $\xi = 0$, only low-quality (LQ) nodes are present, whereas at $\xi = 1$, only high-quality (HQ) nodes exists. In conclusion, the efficiency awareness technique can be used to strategically improve the fidelities of initial paths to manage workload in a quantum network.

\subsubsection{Blocking Probability Study}

\begin{figure}[tb]
    \includegraphics[width=\columnwidth]{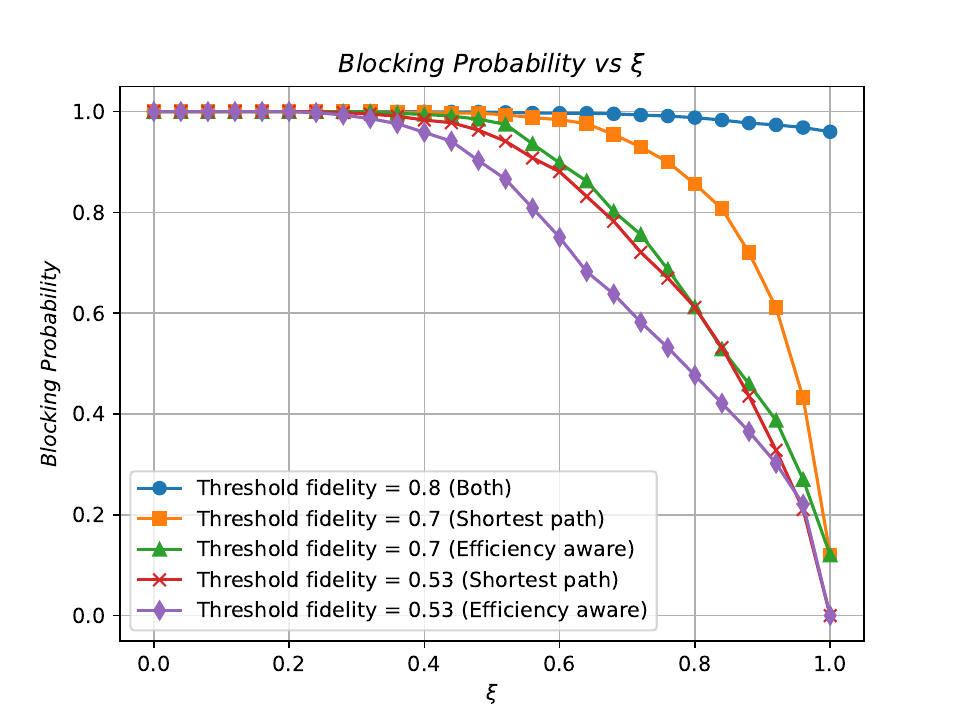}
    \vspace{-0.6cm}
    \caption{Blocking Probability vs. $\xi$}
    \label{fig:bp_vs_hq}
    \vspace{-0.5cm}
\end{figure}
\begin{figure}[tb]
    \includegraphics[width=\columnwidth]{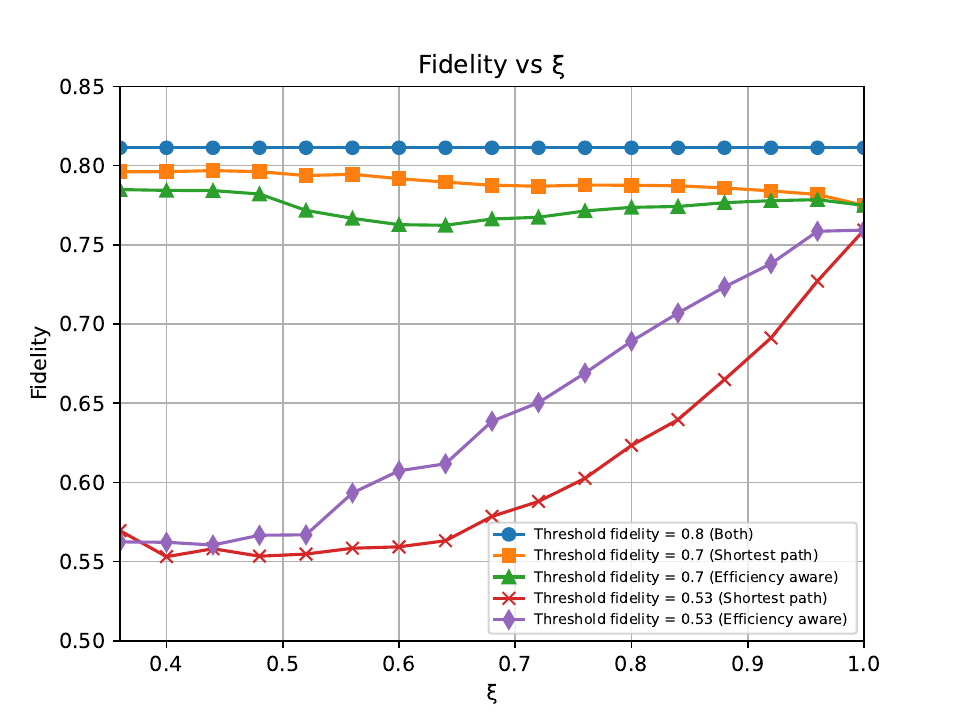}
    \vspace{-0.6cm}
    \caption{Fidelity vs. $\xi$ with fidelity threshold}
    \label{fig:fid_vs_xi_thres}
    \vspace{-0.5cm}
\end{figure}
In this final aspect, we introduce the threshold fidelity $\bar{F}$ into the simulation, which is set to 0.53, 0.7, and 0.8, respectively, in a cylindrical network topology. In addition to the methodology followed in the above study, we reject paths under specific threshold fidelities. 
Note that values below 0.5 are too low for any practical use cases.

Figure~\ref{fig:bp_vs_hq} illustrates the relationship between blocking probability and $\xi$ across various values of $\bar{F}$ with the corresponding fidelity delivered in Figure~\ref{fig:fid_vs_xi_thres}. It is important to highlight that when $\xi < 0.35$, none of the paths achieved fidelity levels exceeding their respective thresholds. Consequently, these paths are designated as blocked and are subsequently excluded from the plot. Notably, a lower $\bar{F}$ corresponds to a reduced blocking probability, as fewer paths fall below the threshold value regardless of the approach employed.

In the comparative analysis depicted in Figure~\ref{fig:bp_vs_hq}, it is evident that the efficiency-aware approach yields a markedly lower blocking probability than shortest-path approach when the fidelity threshold, $\bar{F}$, is set at ${0.53, 0.7}$. As $\bar{F}$ approaches 0.8, however, the disparity between the two methods diminishes. This convergence in performance can be attributed to the high fidelity threshold nearing 0.8, at which point the blocking probability approaches unity. Consequently, at such elevated thresholds, the opportunity for the efficiency-aware approach to further improve performance becomes negligible.  
In conclusion, the efficiency awareness technique improves the network performance in terms of blocking probabilities, i.e., fewer paths would be blocked upon using the efficiency awareness technique.

\section{Conclusions}\label{sec:conclusions}
In this paper, we provide insights into architecting and operating quantum networks with quantum repeaters characterized by different efficiency figures. We showed that incorporating the knowledge of the quality of the nodes into the routing process increases the network performance considerably in terms of fidelity boost to initial paths and lowering of blocked routing paths.  This technique is helpful not only in better network performance but also in motivating better traffic management by assigning requests to path orders depending upon their fidelity threshold. The major bottleneck in the performance of quantum networks lies with the low-quality nodes. Both their number and their efficiency figure play a major role in reduced network performance. For instance, our simulations show that a single replacement of a low-quality node with a high-quality node at the bottleneck point could increase network performance by 50\%.




\section{Acknowledgment}
This work was supported by project NQSTI (PE0000023) under the MUR National Recovery and Resilience Plan funded by the European Union -- NextGenerationEU.


\end{document}